\documentclass{bioinfo}

\usepackage{graphicx}
\usepackage{verbatim}
\usepackage{amsmath}
\usepackage{algorithm}
\usepackage[noend]{algpseudocode}
\usepackage{float}

\newfloat{algorithm}{t}{lop}
\makeatletter
\def\BState{\State\hskip-\ALG@thistlm}
\makeatother
\algnewcommand\algorithmicinput{\textbf{input:}}
\algnewcommand\Input{\item[\algorithmicinput]}
\algnewcommand\algorithmicoutput{\textbf{output:}}
\algnewcommand\Output{\item[\algorithmicoutput]}
\copyrightyear{2015}
\pubyear{2015}

\newcommand{\Eb}{\hat{E}}

\begin{document}
\firstpage{1}

\title[SANA: Simulated Annealing Network Alignment Applied to Biological Networks]{SANA: Simulated Annealing Network Alignment Applied to Biological Networks}
\author[N.Mamano and W.Hayes]{Nil Mamano and Wayne Hayes\footnote{to whom correspondence should be addressed}}
\address{Department of Computer Science, University of California, Irvine CA 92697-3435, USA
}

\history{Received on XXXXX; revised on XXXXX; accepted on XXXXX}

\editor{Associate Editor: XXXXXXX}

\maketitle

\begin{abstract}
The alignment of biological networks has the potential to teach us as much about biology and disease as has sequence alignment.
Sequence alignment can be optimally solved in polynomial time. In contrast, network alignment is $NP$-hard,
meaning optimal solutions are impossible to find, and the quality of found alignments depend strongly upon the algorithm used
to create them.
Every network alignment algorithm consists of two orthogonal components: first, an objective function or measure $M$ that is used to evaluate the quality of any proposed alignment; and second, a search algorithm used to explore the exponentially large set of possible alignments in an effort to find ``good'' ones according to $M$.  Objective functions fall into many categories, including biological measures such as sequence similarity, as well as topological measures like graphlet similarity and edge coverage (possibly weighted). Algorithms to search the space of all possible alignments can be deterministic or stochastic, and many possibilities have been offered over the past decade. In this paper we introduce a new stochastic search algorithm called SANA: Simulated Annealing Network Aligner. We test it on several popular objective functions and demonstrate that it almost universally optimizes each one significantly better than existing search algorithms.
Finally, we compare several topological objective functions using SANA.
Software available at http://sana.ics.uci.edu.
\begin{comment}
Text version of the abstract:

The alignment of biological networks has the potential to teach us as much about biology and disease as has sequence alignment.
Sequence alignment can be optimally solved in polynomial time. In contrast, network alignment is NP-hard, meaning optimal solutions are impossible to find, and the quality of found alignments depend strongly upon the algorithm used to create them.
Every network alignment algorithm consists of two orthogonal components: first, an objective function or measure M that is used to evaluate the quality of any proposed alignment; and second, a search algorithm used to explore the exponentially large set of possible alignments in an effort to find "good" ones according to M.  Objective functions fall into many categories, including biological measures such as sequence similarity, as well as topological measures like graphlet similarity and edge coverage (possibly weighted). Algorithms to search the space of all possible alignments can be deterministic or stochastic, and many possibilities have been offered over the past decade. In this paper we introduce a new stochastic search algorithm called SANA: Simulated Annealing Network Aligner. We test it on several popular objective functions and demonstrate that it almost universally optimizes each one significantly better than existing search algorithms.
Finally, we compare several topological objective functions using SANA.
\end{comment}
\section{Contact:} \href{whayes@uci.edu}{whayes@uci.edu}
\end{abstract}

\section{Introduction}

\subsection{Context}\label{context}

%Introduce network alignment generally
Network alignment is the task of finding the best way to ``fit'' one network inside another.
%It covers a wide range of techniques for finding topologically or functionally similar regions between different networks.
It has applications in several areas, including ontology matching~\citep{DBLP:journals/tkde/LiTLL09}, pattern recognition~\citep{Zaslavskiy:2009:PFA:1638615.1639283}, language processing~\citep{bayati2009-network-alignment}, and social networks~\citep{DBLP:journals/corr/ZhangT13}.
%Network alignment is the problem of finding topologically or functionally similar regions between different networks.
Thus, the specific goal of network alignment depends on the context. We focus on a particular application from the computational biology domain: the alignment of protein-protein interaction (PPI) networks. In a PPI network, nodes represent proteins from a given organism, and edges connect proteins that interact physically. These kinds of interactions are discovered through high throughput experimental methods such as yeast two-hybrid screening~\citep{Ito01022000} or protein
complex purification via mass-spectrometry~\citep{16554755}.
%, thanks to which we dispose of increasingly large amounts of biological network data

% Motivate network alignment of PPI networks
PPI network alignment has many interesting applications. It can serve to transfer biological information across species~\citep{GRAAL}, which, in turn, has been used to offer insights on the mechanisms of human diseases~\citep{bioinflmu-567}, or the process of aging in humans~\citep{Milenkovic:2013:GNA:2506583.2508968}.

% Overview of different types of NA
Network alignment can be classified as local or global. The former aims to align small regions accurately~\citep{PathBlast}. Consequently, it often fails to find large conserved connected subgraphs in different networks. By contrast, global network alignment aims to generate one-to-one node mappings
between two networks. By aligning entire networks, it overcomes the shortcomings of local network alignment. For this reason, the majority of recent research has focused on global network alignment (see section~\ref{state}). There are also methods that allow alignments between more than two networks~\citep{IsoRankN,netcoffee}.

% Explanation of global NA and its goals
We focus on pairwise global network alignment. Its goal is to find a one-to-one mapping from the proteins of the smaller network to proteins in the larger network. Ideally, we would like to find the most biologically relevant mapping: aligned proteins in both networks should be homologically related, in the sense that they used to be {\it same} protein in the species which was the common ancestor of the species of both PPI networks. Since proteins may have multiple descendants, such a mapping may not be one-to-one. For now we ignore this complication and view global pairwise 1-to-1 network alignment as a convenient approximation to the truth.

\subsection{Previous work}\label{state}
% Give a general overview of the biology vs topology issue
Current PPI network alignment methods use a combination of biological and topological information to align similar proteins. Biological information includes \textit{a priori} knowledge about the proteins, such as amino acid sequences. On the other hand, topological information is extracted exclusively from the structure of the PPI network. Since the goal is to obtain biologically relevant alignments, early methods focused on biological information. However, as our understanding of topology-function
relationships~\citep{functionTopologyRelationship} has improved, topology information has gradually shifted to a central role. For instance, it has been shown recently that topological information is more important than sequence information for uncovering functionally conserved interactions~\citep{LGRAAL}.
Topological knowledge can be extracted in many forms. For example, the wiring patterns in the vicinity of homologous proteins in different networks tend to be similar. This information is well captured by graphlets~\citep{Przulj12122004}, which generalize the concept of the degree of a node.

% The hardness of NA motivates heuristic approaches
There are a wide variety of network alignment methods. This diversity is motivated by the inherent computational complexity of network alignment: topologically speaking, one would like to find an alignment that maximizes the number of preserved interactions, i.e., interactions between proteins that are mapped to proteins that also interact. However, this problem is \textit{NP-hard} because it is a generalization of \textit{subgraph isomorphism}, which is \textit{NP-complete}~\citep{Cook:1971:CTP:800157.805047}. This means that no efficient algorithm is known. Thus, practical methods must rely on approximation and heuristic techniques; and when it comes to heuristic algorithms, the possibilities are endless but there is no obvious ``best option''.

% mention previous methods
In the biological network domain, work over the past few years has included IsoRank~\citep{Isorank}, the family of GRAAL algorithms (GRAAL~\citep{GRAAL}, H-GRAAL~\citep{HGRAAL}, C-GRAAL~\citep{CGRAAL}, MI-GRAAL~\citep{MIGRAAL}, L-GRAAL~\citep{LGRAAL}), NATALIE~\citep{NATALIE}, GHOST~\citep{GHOST}, NETAL~\citep{NETAL}, SPINAL~\citep{SPINAL}, PISwap~\citep{PISwap} MAGNA~\citep{MAGNA} and its successor MAGNA++~\citep{MAGNApp},  HubAlign~\citep{HubAlign}, and OptNetalign~\citep{optnetalign}. \cite{survey2015} did a survey of existing methods, datasets, and optimization measures.

% distinguish between obj functions and search algorithms
% local vs global measures
In general, each method defines an \textit{objective function} or \textit{measure} over alignments that can be viewed as a \textit{score}, and then proposes a \textit{search algorithm} that searches through the enormous space of possible alignments in an attempt to maximize the objective function. Some measures such as sequence similarity %there is no refence in previous papers for sequence similarity
or graphlet similarity~\citep{GRAAL} are defined over pairs of proteins instead of whole alignments. They can be generalized to whole alignments by taking the average similarity score among all pairs of aligned proteins. These measures are called \textit{local measures}, because the contribution of each mapping is independent of the others. On the other hand, global measures aim to evaluate the alignment from a global perspective, and not on a node-to-node basis.

% discussion about search algorithms for local measures
% not essential
When the objective function is a local measure, the search algorithms used to maximize it usually fall in three categories: greedy best-first algorithms~\citep{NETAL,GREAT}, seed-and-extend algorithms~\citep{GRAAL,MIGRAAL,SPINAL,HubAlign}, and the Hungarian algorithm~\citep{HGRAAL,GREAT}. A greedy best-first algorithm starts with an empty alignment and progressively aligns the most similar pair among the available pairs until it is complete. A seed-and-extend algorithm usually starts aligning the most similar node pair (called the seed pair), and then proceeds in a local fashion by aligning the most similar pair among their neighbors. The local nature of seed-and-extend algorithms usually results in large common connected subgraphs. Finally, the Hungarian algorithm can solve the problem optimally as long as the measure optimized is only local. However, optimality in the local objective function does not imply optimality in other measures.
%However, since the alignments' quality is finally evaluated according to a different set of measures (see section~\ref{eval}), this does not necessarily mean that this method is superior to the others.

% distinguish between obj functions and target measures
While every method has its own objective function, the alignments are compared according to a set of \textit{target} measures that have been established as the most important (see section~\ref{eval}). However, even target measures are of heuristic nature, because except when we align a network with itself, the \textit{correct} mapping is unknown. A good objective function should guide the search algorithm to alignments that score well in all the target measures. Sometimes the objective function can be one of the target measures~\citep{MAGNA}. However, since target measures are usually global measures, the search algorithms for local measures described above are in general not applicable.

% introduce L-GRAAL and HubAlign
% not essential / can be moved to the "Results and discussion" section
Among all the methods for aligning PPI networks, currently L-GRAAL \citep{LGRAAL} and HubAlign \citep{HubAlign} seem to yield the best results. All the methods previously mentioned have been shown to be inferior to at least one of them, either directly or indirectly. Both the L-GRAAL and HubAlign objective functions combine a topological measure and sequence similarity. In the case of L-GRAAL, the topological measure is based on graphlets, while its search algorithm uses integer programming. In the case of HubAlign, the topological measure is a local measure that gives higher scores to pairs of topologically important proteins such as those that act as hubs (proteins that have many connections) or bottlenecks. Its search algorithm is a variation of seed-and-extend with multiple seeds.

\subsection{Our contributions}

We present SANA (\textit{Simulated Annealing Network Aligner}), a search algorithm based on Simulated Annealing~\citep{SA1,SA2}, a metaheuristic local search algorithm with a rich history of successful applications to many optimization problems across a wide variety of domains.
%and show that it is effective, efficient, simple, and flexible.
It shares many characteristics with MAGNA~\citep{MAGNA}, a genetic algorithm. Both can be used to optimize any objective function directly, and both can start with any alignment and then improve it. However, SANA converges to a good solution faster than most existing algorithms.
It is conceptually simple to implement, although it has a few user-defined parameters that need to be optimized to get the best results. This can be done either manually by trial-and-error, or automatically~\citep{autosa}.

\section{Methodology}
\begin{methods}
\subsection{Main idea}

%intuition of annealing
Annealing is a process used in metallurgy to create crystals. A crystal is a highly structured, low-energy state of a material that can only be reached if the material is cooled at a very specific rate. If the substance cools at the ``correct'' rate, the material can correctly ``find'' the lowest energy state, which forms the most perfect crystal.  It is also similar to a process we're all familiar with: ``shaking'' a box full of small objects in order to have those objects ``settle'' in the box. The settled state has lower energy, and is reached by the random motions caused by shaking the box. 

% analogy between annealing and simulated annealing
Simulated annealing is a metaheuristic algorithm, which means that it is not tailored to any specific optimization problem. It can be applied to any optimization problem as long as the necessary elements are defined: the \textit{solutions}, the \textit{objective function} and the \textit{neighbor relationship}. 
The analogy to annealing goes as follows: A solution is like an state of the material. In our case, a solution is an alignment. The objective function is analogous to the energy of the material. While in metallurgy the goal is a state with minimum energy, in simulated annealing the best solution minimizes (or maximizes) some arbitrary objective function of our choice.
When atoms move due to high temperature, the state of the material changes slightly. In order to simulate this, we need a neighbor relationship that indicates which solutions are close to each other. Then, we can change a solution for a neighbor solution. For instance, we can say two alignments are neighbors if they only vary in one or two mappings of individual pairs of aligned nodes.

% intuition behind simulated annealing
If we take a random alignment, called the \textit{initial solution}, we can improve it by looking at its neighbor solutions and choosing the best one. If we repeat this process, we will reach a \textit{local minimum} quickly in which we can no longer improve. However, since the energy landscape is unlikely to be monotonic everywhere, this local minimum is unlikely to be the global minimum. To avoid this pitfall, simulated annealing introduces the ability to allow worse solutions to be selected with some probability, analogous to how high-temperature materials have enough energy to move freely through different states. As the temperature decreases, the ability to escape local minima decreases. If the temperature schedule is chosen correctly, then the solution will tend towards a global minimum.

\subsection{SANA algorithm}
% definitions and notations
Let $G_1=(V_1,E_1)$ and $G_2=(V_2,E_2)$ be two graphs (networks) with $|V_1|\leq|V_2|$. A pairwise global alignment $a$ (from this point simply alignment) from $G_1$ to $G_2$ is an injective function from $V_1$ to $V_2$. An objective function $f$ is a function from the set of all alignments to the closed range $[0,1]$.

% algorithm
The basic scheme of SANA is shown in Algorithm~\ref{alg:SANA}. The input consists of the two networks $G_1,G_2$, an objective function $f$, a starting alignment (in the absence of one, SANA generates a random alignment), and a maximum execution time $t_{max}$. Since SANA is a generic search algorithm, $f$ can be any objective function. The output is an alignment $a$ that aims to optimize $f$---in our case, we {\it maximize} various topological and biological similarity measures rather than minimizing energy.

An important element of simulated annealing is the \textit{temperature schedule}, which determines the decline of the probability to accept a worse solution as the algorithm advances. The \textit{temperature} $T(i)$ is a control parameter which depends on the current iteration $i$. It is commonly defined as $T(i)=k\cdot e^{-\lambda\cdot i}$, where $k$ and $\lambda$ are empirically determined constants greater than zero~\citep{SA1}. The temperature is at its highest point at iteration 0 where $T(0)=k$, and approaches 0 asymptotically. The constant $\lambda$ determines how fast the temperature approaches 0.

We define two types of neighbors among alignments: \textit{change} and \textit{swap} neighbors (see Figure~\ref{fig:operators}). Change neighbors differ only in one mapping, which has the same origin in $G_1$ but different destinations in $G_2$. Swap neighbors differ in exactly two mappings, which have the same sources but their images are exchanged. Together, the two types of neighbors allow SANA to explore the solution space completely: through a sequence of neighbors, it is possible to go from any alignment to any other alignment. In SANA when a random neighbor is generated (line 3 of Algorithm~\ref{alg:SANA}) the probability of choosing each type of neighbor is proportional to its \textit{branching factor}, i.e., the number of different neighbors of $a$ of that type. This way, all neighbors are equally likely. The idea of using swaps to improve the alignment is not new; it has been used before with other local search algorithms~\citep{PISwap,MAGNA}.

In the algorithm, $\Delta E$ denotes the energy increment between the new and the current solution. The probability to accept a worse solution is $P(\Delta E, T(i)) = e^{\Delta E/T(i)}$. This probability decreases when the difference between $f(a)$ and $f(a')$ increases, and it also decreases as the temperature decreases.
%We implement the fact that the neighbor is accepted with probability $P(\Delta E, T(i))$ by comparing this value with a uniformly distributed pseudorandom number in the range $(0,1)$ (represented by $\mbox{random}(0,1)$ in Algorithm~\ref{alg:SANA}) and only accepting the new solution if $P(\Delta E, T(i))$ is higher.

\begin{algorithm}
\caption{SANA}\label{alg:SANA}
\begin{algorithmic}[1]
\Input{$G_1,G_2,f,a_0,t_{max}$}
\Output{$a$}
\State Let $a=a_0, i=0$
\While {$t_{exec}<t_{max}$}
\State $a'\gets \mbox{random neighbor(a)}$
\State $\Delta E \gets f(a')-f(a)$
\If {$\Delta E \geq 0$} $a\gets a'$
\Else \State $a\gets a'$ with probability $P(\Delta E, T(i))$
\EndIf
\State $i \gets i+1$
\EndWhile
\State \Return $a$
\end{algorithmic}
\end{algorithm}

%visualize neighbors
\begin{figure}
\centering
\includegraphics[width=0.7\linewidth]{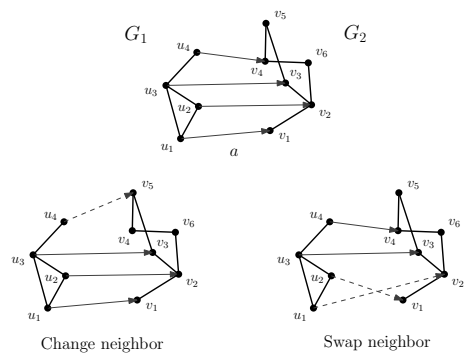}
\caption{{\bf Top}: an alignment between $G_1$ on the left and $G_2$ on the right, with the alignment depicted by horizontal arrows.
{\bf Bottom left}: a {\it change} neighbor: the alignment of node $u_4$ has moved from $v_4$ to $v_5$.
{\bf Bottom right}: a {\it swap} neighbor: the alignments $(u_1,v_1)$ and $(u_2,v_2)$ have swapped to be $(u_1,v_2)$ and $(u_2,v_1)$.}
\label{fig:operators}
\end{figure}

As shown in Figure~\ref{fig:parameters}, with appropriate values of $k$ and $\lambda$ the temperature schedule allows the algorithm to move freely through the solution space at the beginning, gradually becoming more selective until it stagnates at a local maximum.

% discussion about the effect of k and lambda
\begin{figure}
\centering
\includegraphics[width=0.99\linewidth]{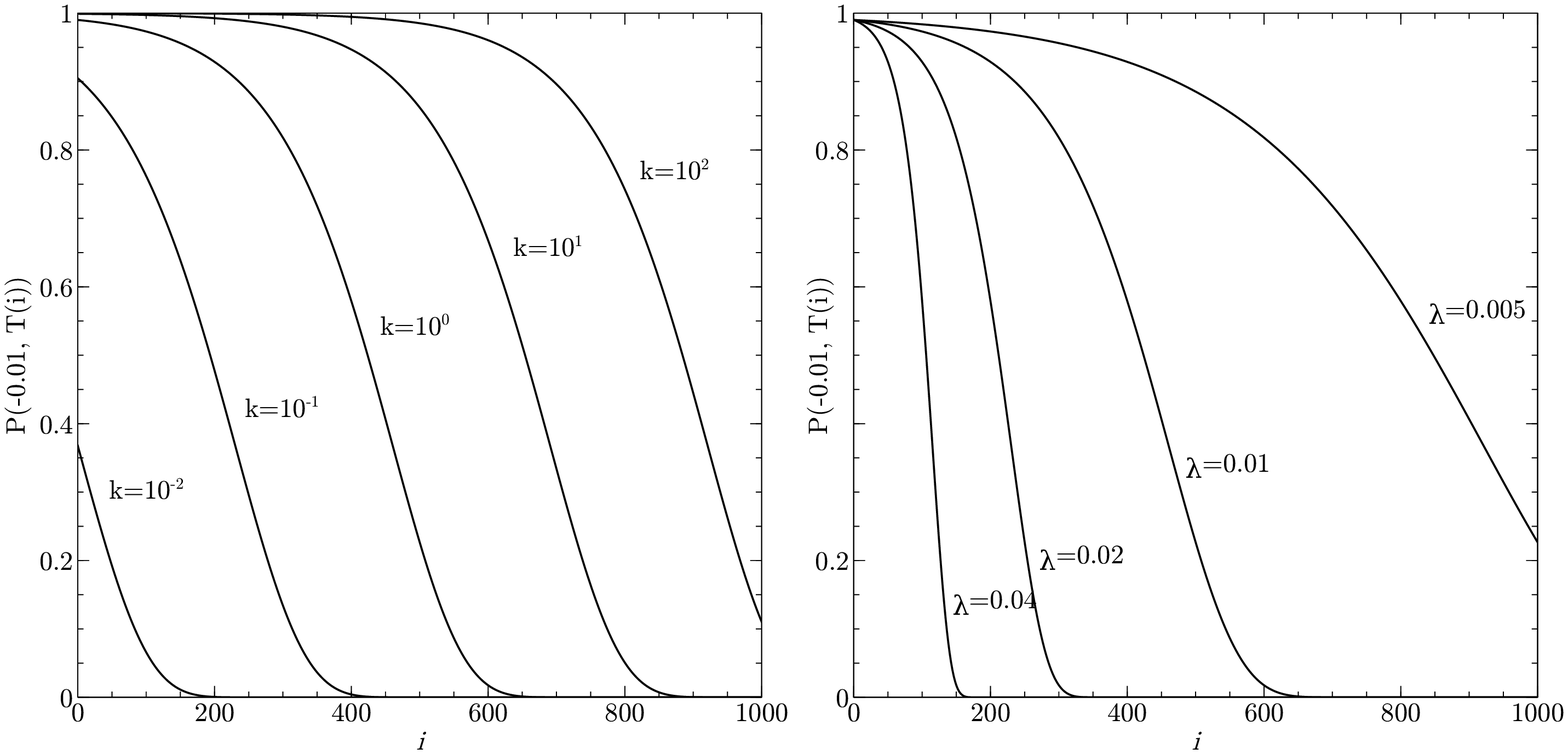}
\caption{Effect of the $k$ and $\lambda$ constants in the temperature schedule. The figures show the probability to adopt a worse solution with an energy increment $\Delta E = -0.01$ as a function of the iteration $i$. In the left figure, $\lambda$ is fixed to $\lambda=0.01$. In the right figure $k$ is fixed to $k=1$.}
\label{fig:parameters}
\end{figure}

%Despite the capacity of simulated annealing to escape local maximums, the final result is still greatly conditioned by the initial stages of the process. Some initial starting points may be too far from any good solution, in the sense that it would take far too many individual changes or swaps (see Figure~\ref{fig:operators}) to get from the initial point to a good solution.
%Thus, in the absense of a starting alignment, SANA works in threes stages: (i) generate a large number of random starting alignments and run all of them for a short time; (ii) keep the best $K$, and run these a bit further; and (iii) choose the best of the $K$ to run to completion.  For now we tune these three stages by hand.

\subsection{Incremental evaluation}\label{implementation}

A SANA iteration consists of updating the temperature, generating a neighbor alignment, evaluating it, and deciding whether to keep it. It is clear that all the steps other than evaluating the new alignment are constant time operations. Thus, the step that will determine the running time of each iteration is the alignment evaluation. Depending on the objective function, evaluating an alignment may require visiting every node in $G_1$ and $G_2$. However, between two neighbor alignments only one or two mappings are affected. In this section we show that for many typical objective functions, we can efficiently compute the score of the neighbor alignment incrementally from the score of the original alignment.

\emph{\textbf{Edge Coverage\footnote{Often called ``Edge Correctness'', although coverage is a better term because for most alignments no meaningful definition of ``correct'' exists. See Appendix.} (EC)}} is the fraction of edges in $G_1$ that are aligned to ({\it i.e.} cover) edges in $G_2$. Coverage represents interactions between proteins in one network that are mapped to proteins that also interact in the other, although there is no requirement that the mapped interactions are in any sense biologically equivalent. Let $E_a=\{(u_1,v_1)\in E_1\mid (a(u_1),a(v_1))\in E_2\}$ denote the set of edges in $G_1$ that cover edges in $G_2$ in alignment $a$. Then, the edge coverage of $a$ is $EC(a)={|E_a|}/{|E_1|}$. 

Evaluating whether an edge is covered requires constant time, so EC can be computed in time $O(|E_1|)$ by checking if each edge in $E_1$ belongs to $E_a$. In order to compute EC incrementally, we need to see how the set of covered edges changes between neighbor solutions. Assume that we know $E_a$ for some alignment $a$ and we need to know $E_{a'}$, where $a'$ is change neighbor of $a$. Let $u_1$ be the node in $G_1$ such that $a(u_1)\not= a'(u_1)$. Note that for any edge not incident to $u_1$, the image of its two endpoints has not changed, and thus whether it is covered by an edge in $G_2$ has not changed either. This means that we only need to check the edges incident to $u_1$. Therefore, the cost of computing $E_{a'}$ incrementally is $O(\mbox{degree}(u_1))$. The amortized cost of finding $E_{a'}$ for change neighbors is $O(|E_1|/|V_1|)$, the average degree of nodes in $G_1$. The case of swap neighbors is analogous, as we only need to consider the edges incident to the two swapped nodes. For sparse networks such as PPI networks, $|E_1|$ is typically within a small constant factor of $|V_1|$ (see table~\ref{networkTable}) so the practical amortized cost is constant.

\emph{\textbf{Symmetric Substructure Score (S3)}}~\citep{MAGNA} A drawback of edge coverage is that it does not penalize an alignment for mapping sparse regions of $G_1$ to dense regions in $G_2$. S3 corrects for this by penalizing the mapping of interactions in $G_1$ to non-interactions in $G_2$. Let $\Eb_a=\{(u_2,v_2)\in E_2 \mid \exists u_1,v_1 \in V_1 \wedge a(u_1)=u_2 \wedge a(v_1)=v_2\}$ denote the set of edges of the subgraph of $G_2$ induced by the nodes in the alignment $a$. Then, S3 is defined as: $S3(a)=\frac{|E_a|}{|E_1|+|\Eb_a|-|E_a|}$.

We have seen how to compute $E_{a'}$ incrementally. The set $\Eb_{a'}$ can also be computed similarly. Between change neighbors, if for instance $a(u_1)=u_2$ and $a'(u_1)=v_2$, all the edges incident to $u_2$ will not be in $\Eb_{a'}$, and the edges incident to $v_2$ may be in it. Between swap neighbors $a$ and $a''$, we have $\Eb_{a}=\Eb_{a''}$. Hence, the cost of computing $\Eb_{a'}$ incrementally is $O(\mbox{degree}(v)+\mbox{degree}(w))$, and the amortized cost is $O(|E_2|/|V_2|)$.

\emph{\textbf{Local measures}}. Given some notion of similarity between proteins, such as sequence or graphlet similarity, let $s(u,v)$ denote the similarity between two nodes $u\in V_1, v\in V_2$. A local measure
$M_s(a)=\sum_{u\in V_1} s(u,a(u))/|V_1|$
is the average similarity of the aligned proteins.
We can compute the score difference of a local measure between neighbor alignments by subtracting the similarity of the node pairs that are no longer aligned and adding the similarity of the newly aligned pairs. If we assume that all the required similarities $s(u,v)$ are known before SANA starts, this requires constant time, as neighbor alignments only vary in one or two mappings.

\emph{\textbf{Weighted Edge Coverage (WEC)}}~\citep{WAVE} generalizes EC by making the contribution of each covered edge not equal, but a function of the similarity of the aligned endpoints. In order to define the similarity of the nodes, we can use any local measure $s$. Then, for each edge $(u_1,v_1)\in E_1$ covered by the alignment, its contribution is $\left(s(u_1,a(u_1))+s(v_1,a(v_1))\right)/2$. The WEC score is
$$\frac{1}{|E_1|}\sum_{(u_1,v_1)\in E_a} \left(s(u_1,a(u_1))+s(v_1,a(v_1))\right)/2$$
The WEC measure can be optimized incrementally by adapting the method used for EC to account for the weights.

\emph{\textbf{Combinations}} Any weighted combination of the aforementioned objective functions can also be computed efficiently by computing each one incrementally and then multiplying their scores by the corresponding factor to obtain the final score.

Incremental evaluation makes SANA scalable to enormous PPI networks, as the cost of each step only depends on the average degree of nodes. For objective functions that cannot be evaluated incrementally, SANA may become prohibitively expensive.

\end{methods}

\section{Results}

\subsection{Alignment evaluation}\label{eval}
In this section we describe the target measures we use to assess the quality of alignments. These measures are the same used by most previous methods. 

For {\em topological measures}, we use Edge Coverage (EC), Symmetric Substructure Score (S3), Node Correctness (NC) and Largest Common Connected Subgraph (LCCS).  The first two are described in Section~\ref{implementation}.

Assuming there exists a correct alignment between $G_1$ and $G_2$, the {\it Node Correctness (NC)} is the fraction of correctly aligned nodes. This is the most important measure, but since in general the correct alignment between different species is not known, it can only be used when aligning a network with itself.

The \textit{common subgraph} of an alignment $a$ between $G_1$ and $G_2$ is the subgraph of $G_1$ that remains when considering only edges covered by the alignment: $CS_a=(V_1,E_a)$. A good alignment has a common subgraph with large connected regions. Let $CC_l=(V_l,E_l)$ be the largest connected component of $CS_a$. LCCS measures the size of $CC_l$ as the geometric mean of (i) the fraction of nodes in $CC_l$, $|V_l|/|V_1|$, and (ii) the fraction of edges in $CC_l$, $|E_l|/\min(|E_1|,|\Eb_a|)$~(Kuchaiev et al., 2010; more detail in Saraph and Milenkovi\'{c}, 2013).

To assess {\it biological quality}, we measure common Gene Ontology (GO) terms~\citep{GOterms}. More precisely, we are interested in the fraction of aligned proteins sharing at least $k$ GO terms, for $k=1\ldots 9$. We denote this local measure $\mbox{GO}_k$. Some GO terms are very common, and thus aligning them does not have much significance. To account for this fact, for each network we removed the most common terms until no more than half of the total GO occurrences remained. Although ostensibly more sophisticated measures using GO terms exist, none has yet become dominant and we believe this simple, easy-to-interpret GO-based measure is the one of the best available at the moment.

\subsection{Datasets}\label{datadescription}

We attempt a comprehensive analysis by using three different datasets used by several previous studies.

\emph{\textbf{Noisy yeast}} This dataset consists of six variations of a yeast PPI network~\citep{Collins01032007}. They all have the same set of nodes and vary only in the number of edges. The first network contains 8,323 interactions, and the other five are formed by progressively increasing the number of edges by 5, 10, 15, 20, and 25\% with lower-confidence interactions from the same experiment. We align the first network against itself and each noisy variant. Since the underlying network is the same, we know the true node mapping. This dataset has been used by many previous authors~\citep{MAGNA,GHOST,WAVE,GREAT}.

\emph{\textbf{BioGRID dataset}} This dataset includes eight networks from the manually curated BioGRID database~\citep{Chatr-aryamontri01012013} (v3.2.101, June 2013). Among the six networks with lowest amount of edges, we align each pair of networks. We also align the two networks with more edges (SC and HS). This dataset was used in~\cite{LGRAAL}.

\emph{\textbf{Yeast and human}} This dataset consists of two older Yeast and Human networks~\citep{Collins01032007,humanNetwork}. This is one of the most often used datasets. It was used in~\citep{GRAAL,NETAL,MAGNA,MIGRAAL}. 

Note that Yeast and Human have more than one dataset associated with them, for consistent comparison with the results of previous authors.
\begin{table}[h]
\label{networkTable}
\caption{List of networks grouped by datasets, with the corresponding identifiers used in the plots.}
\begin{tabular}{llrr}
\hline
\textbf{Network}                   & \textbf{Identifier} & \multicolumn{1}{l}{\textbf{Proteins}} & \multicolumn{1}{l}{\textbf{Edges}} \\ \hline
Yeast                              & Y0                  & 1,004                                 & 8,323                                     \\
Yeast (+5\% noise)                 & Y5                  & 1,004                                 & 8,739                                     \\
Yeast (+10\% noise)                & Y10                 & 1,004                                 & 9,155                                     \\
Yeast (+15\% noise)                & Y15                 & 1,004                                 & 9,571                                     \\
Yeast (+20\% noise)                & Y20                 & 1,004                                 & 9,987                                     \\
Yeast (+25\% noise)                & Y25                 & 1,004                                 & 10,403                                    \\ \hline
\textit{Rattus norvegicus}         & RN                  & 1,657                                 & 2,330                                     \\
\textit{Schizosaccharomyces pombe} & SP                  & 1,911                                 & 4,711                                     \\
\textit{Caenorhabditis elegans}    & CE                  & 3,134                                 & 5,428                                     \\
\textit{Mus musculus}              & MM                  & 4,370                                 & 9,116                                     \\
\textit{Arabidopsis thaliana}      & AT                  & 5,897                                 & 13,381                                    \\
\textit{Drosophila melanogaster}   & DM                  & 7,937                                 & 34,753                                    \\
\textit{Saccharomyces cerevisiae}      & SC                  & 5,831                                 & 77,149                                    \\
\textit{Homo Sapiens}   & HS                  & 13,276                                 & 110,528                                   \\ \hline
Yeast                              & Y                   & 2,390                                 & 16,127                                    \\
Human                              & H                   & 9,141                                 & 41,456                                    \\ \hline
\end{tabular}
\end{table}

\subsection{Compared methods and parameters}

In order to keep our analysis brief and relevant we compare SANA only against the two currently best-performing methods: L-GRAAL~\citep{LGRAAL} and HubAlign~\citep{HubAlign}. \cite{LGRAAL} performed a thorough analysis demonstrating that for every target measure of interest, either L-GRAAL or HubAlign beat all other methods among GHOST, MI-GRAAL, SPINAL, NETAL, NATALIE, MAGNA, IsoRank, and PISwap (in the BioGRID dataset). Thus, we compare only against HubAlign and L-GRAAL.

We set a maximum execution time of one hour for all methods. In L-GRAAL, we set a limit of 1,000 iterations as in~\citep{LGRAAL}. We use default values for the remaining parameters of L-GRAAL and the parameters of HubAlign.

For SANA, we need to set the constants $k$ and $\lambda$ of the temperature schedule, the optimal values for which change on a case by case basis depending on variables such as the network sizes or the objective function being evaluated. We use an automatized method to find suitable values automatically in a few minutes. It mimics what humans would do by trial-and-error. As initial temperature $k$, it searches for the lowest temperature such that the behavior at this temperature is still random. For the rate of decay $\lambda$, it searches for the slowest decay such that after $t_{max}$ time the temperature is practically zero. The details of this method are out of scope for this paper.

The objective function of SANA, HubAlign, and L-GRAAL is a combination of sequence similarity and a topological measure. A parameter $\alpha$ controls the weight of sequence similarity as opposed to topological similarity, with $\alpha=1$ using only sequence, and $\alpha=0$ using only topology:
\begin{equation}
\mbox{score}(a) = (1-\alpha)T(a) + \alpha S(a)\label{eq:alpha}
\end{equation}
where $T(a)$ and $S(a)$ are the topological and sequence similarity of an alignment $a$, respectively. For sequence similarity, we use normalized BLAST bit-scores~\citep{blast}. If $s(u_1,u_2)$ is the BLAST bit-score of proteins $u_1$ and $u_2$,
$$S(a)=\sum_{u_1\in V_1} \frac{s(u_1,a(u_1))}{\mbox{max}_{v_1\in V_1,v_2\in V_2}s(v_1,v_2)}$$

For the topological measure of SANA, we use one of the following: EC, S3, or the objective function of L-GRAAL, which is WEC with graphlet similarity as local measure~\citep{LGRAAL}.

Our results suggest that L-GRAAL's objective function provides the best all-round measure of topology for alignment purposes. Moreover, using the same objective function as L-GRAAL illustrates how SANA is the key component in finding the best alignments when combined with any objective function. We denote the resulting method SANA-LG, whereas the variants in which the topological component of the objective function are S3 and EC are denoted SANA-S3 and SANA-EC, respectively.

In the plots, we show the average score of three runs (in Figure~\ref{fig:syntheticyeast}) or two runs (in the remaining Figures). Moreover, to ensure fairness (specially when aligning a network to itself in the Noisy Yeast dataset), we randomly shuffle the ordering of the nodes and edges in the input network files before each run. Due to this, even deterministic methods such as HubAlign obtain a different score every time.

\subsection{Topology only comparison}\label{top}

\begin{figure}
\centering
\includegraphics[width=0.99\linewidth]{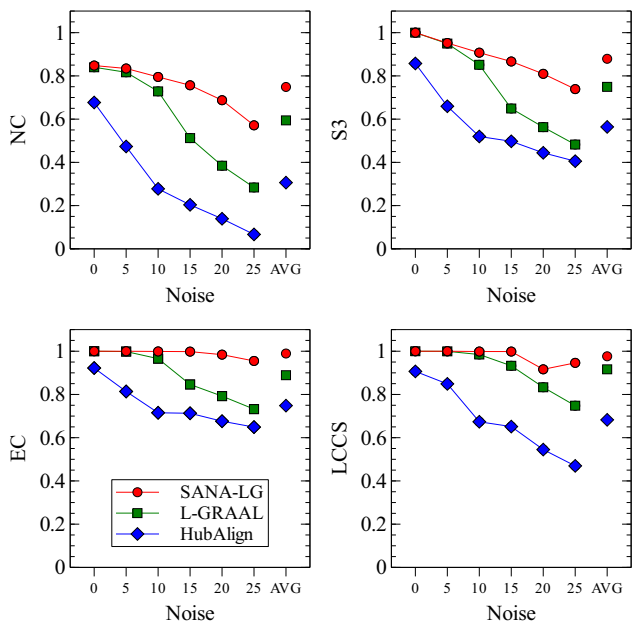}
\caption{Topological measure scores for the Noisy Yeast dataset when using $\alpha=0$. $G_1$ is always Y0 (no noise), and the $x$-axis corresponds to the amount of noise in $\%$ in $G_2$. For SANA-EC we only show the averages for clarity.}
\label{fig:syntheticyeast}
\end{figure}

In this section we concentrate on optimizing the topological quality of our alignments without regard to biology (thus $\alpha=0$).
In Figure~\ref{fig:syntheticyeast} we see the topological scores obtained with the different methods in the Noisy Yeast dataset, where the true node mapping is known. SANA-LG shows more tolerance to noise, as its scores do not decline as quickly as we add noise. This is important because PPI networks tend to have high levels of noise, as the methods used to discover the interactions have an inherent high error rate.

Note that when $G_1$ and $G_2$ have the same number of nodes, optimizing EC and S3 is equivalent. This is because all the nodes in $G_2$ are part of the alignment, and therefore $\Eb_a$, the set of edges in the subgraph of $G_2$ induced by the nodes in the alignment, is simply $E_2$ regardless of the alignment $a$. The only term in the formula of $S3$ that changes as a function of the alignment is $|E_a|$, the number of covered edges, just as in the formula of $EC$. Therefore, for this comparison we omit SANA-S3.

\begin{figure}
\centering
\includegraphics[width=0.99\linewidth]{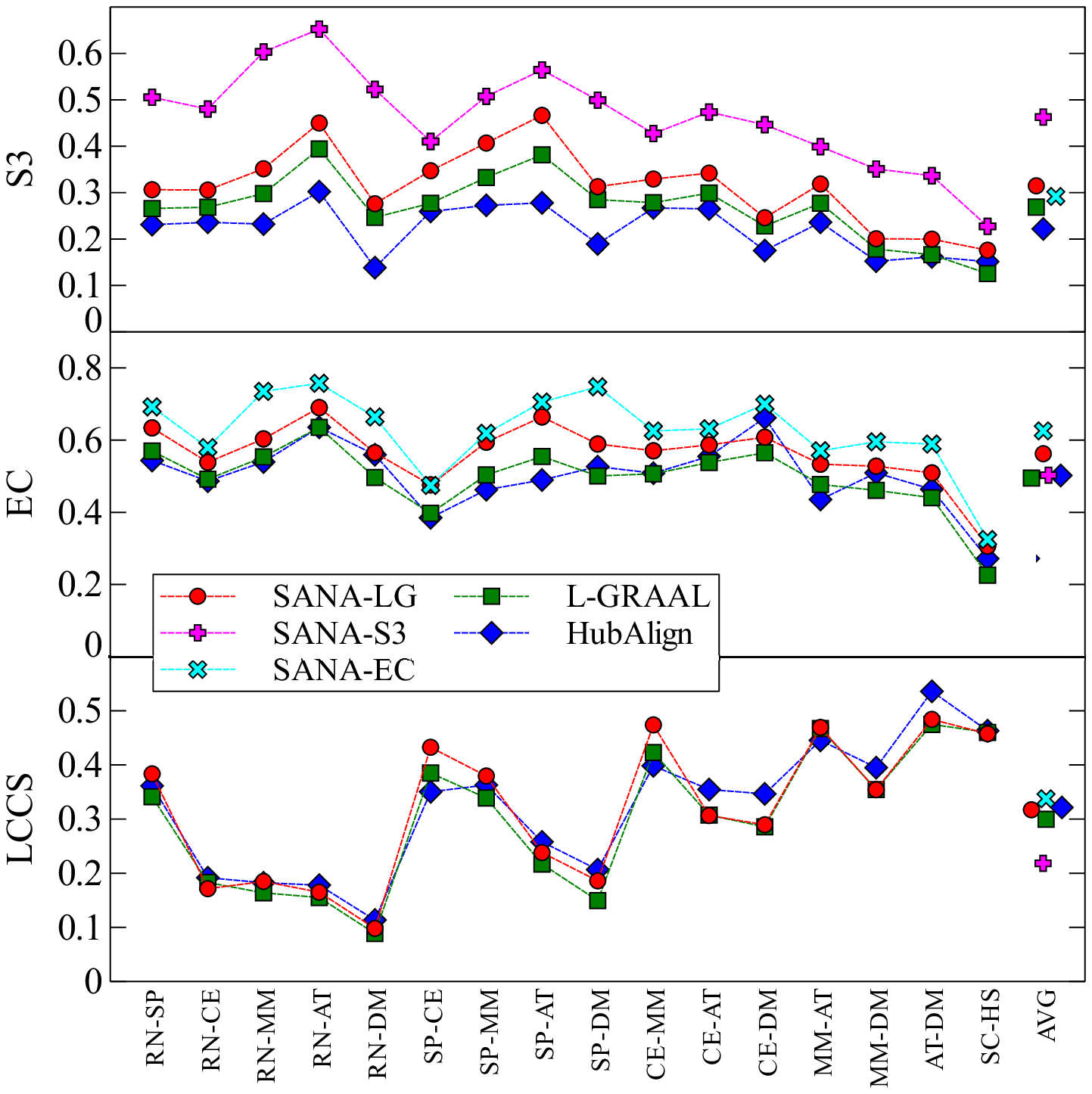}
\caption{Topological measure scores for the BioGRID dataset when using $\alpha=0$. Note that the networks are not sorted in any particular order, so the lines are only a visual aid. The top plot shows that SANA-S3 is by far the best method for optimizing S3; similarly, the middle plot demonstrates that SANA-EC is the best at optimizing EC. For SANA-EC and SANA-S3 we only show the averages for clarity, except in their respective objective functions.}
\label{fig:biogrid_alpha0}
\end{figure}

In Figure~\ref{fig:biogrid_alpha0} we see the topological scores of the BioGRID dataset. SANA-LG performs consistently better than L-GRAAL and HubAlign in all measures except in LCCS, and is only beaten in S3 by SANA-S3 (which explicitly optimizes S3) and is only beaten in EC by SANA-EC (which explicitly optimizes EC).

\begin{figure}
\centering
\includegraphics[width=0.99\linewidth]{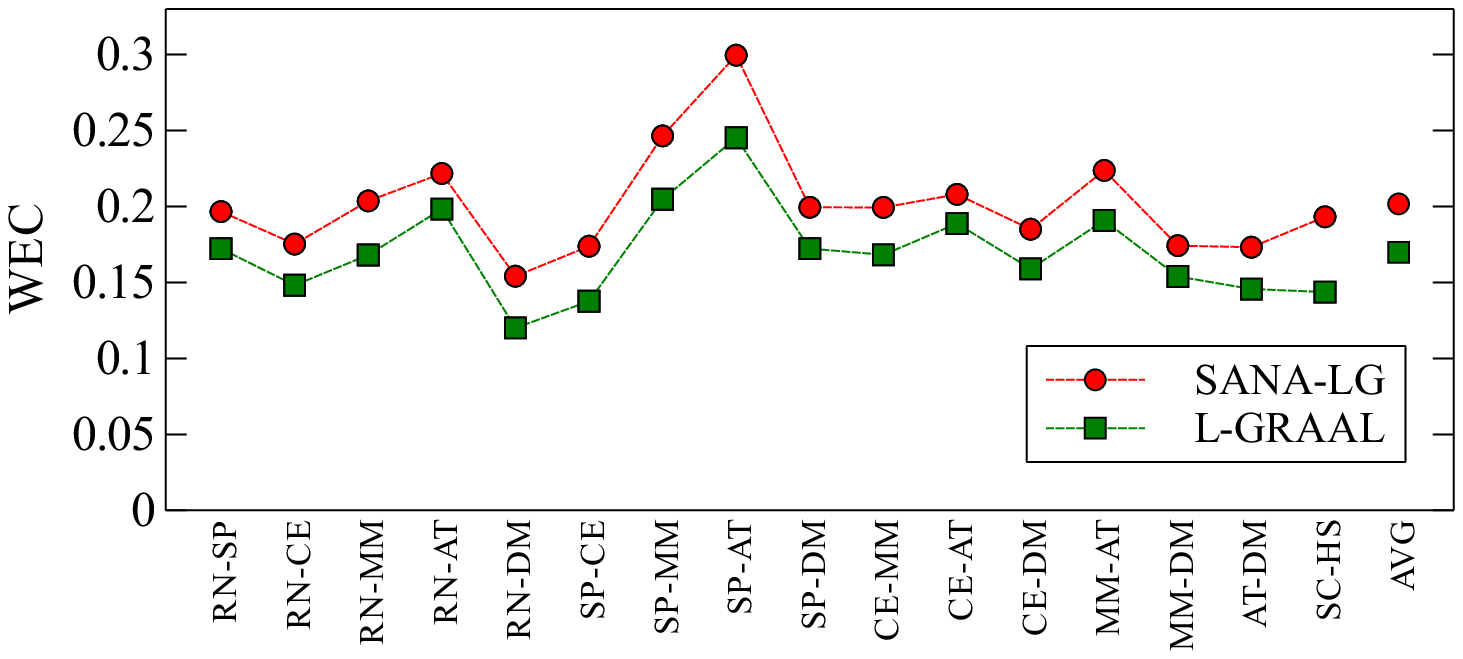}
\caption{Comparison between L-GRAAL alignment strategy and SANA at optimizing L-GRAAL objective function (graphlet-based WEC) when using $\alpha=0$. A similar graph appears if we compare SANA-S3 to MAGNA (which optimizes S3).}
\label{fig:lgraalobjfuncomparison}
\end{figure}

In Figure~\ref{fig:lgraalobjfuncomparison} we compare L-GRAAL's alignment strategy and SANA directly by showing the scores in their common objective function, graphlet-based WEC (with $\alpha=0$), and see that SANA is capable of optimizing L-GRAAL's objective function better than L-GRAAL itself. We can make similar comparisons with other methods. For instance, in the Yeast and Human dataset, MAGNA, which optimizes S3, obtains a S3 score of ${\sim}26\%$ (or ${\sim}20\%$ when starting with a random alignment) while SANA-S3 obtains scores of ${\sim}40\%$ when starting with a random alignment.
% A similar comparison optimizing HubAlign's topological importance-based measure is not possible because in HubAlign the objective function and the search algorithm complement each other: edge coverage is only enforced through the search algorithm. Thus, SANA does not obtain good alignments when optimizing only its objective function.

\subsection{Biological comparison}\label{bio}

In this section we evaluate the methods' ability to find alignments that score well in both topology and sequence. It has previously been shown that using sequence information alone to align pairs of nodes in different species produces alignments that have very low topological quality (comparable to random alignments), but that using graphlet-based topological information alone can recover significant biological information~\citep{GRAAL}. Since then it has also been shown that using both types of information can produce alignments that score reasonably well in both the topological and biological sense. However, balancing the two is not as simple as setting $\alpha=0.5$ in Equation~\ref{eq:alpha} if the topological measure and sequence have different plausible ranges of scores. For instance, when SANA optimizes EC aligning \textit{RNorvegicus} and \textit{SPombe}, it obtains a EC score of 0.69 (Figure~\ref{fig:biogrid_alpha0}). However, when SANA optimizes sequence on the same pair of networks, it obtains a Sequence score of 0.057 (Figure~\ref{fig:biogrid_beta0,5}). This means that using $0.5T(a) + 0.5S(a)$ does not ``balance'' the two, because sequence scores are under-weighted due to their low value; a relatively small increase in EC would compensate for a large drop in Sequence.

For this reason, we introduce a new balancing parameter called $\beta$ that accounts for differences between expected topological and sequence scores. For instance, $\beta=0.5$ means that topology and sequence are optimized at equal parts. Given a certain $\beta$, we can compute the $\alpha$ that achieves the corresponding balance as
$$\alpha = \frac{\beta\cdot S_{top}}{(1-\beta)\cdot S_{seq}+\beta\cdot S_{top}}$$
where $S_{top}$ is the score of the method using $\alpha=0$, and $S_{seq}$ is the score of the method using $\alpha=1$, i.e.\ optimizing only sequence\footnote{HubAlign does not work properly when optimizing Sequence exclusively, so for this method we used $\alpha=0.9999$ to compute $S_{seq}$.}. Since $S_{top}$ and $S_{seq}$ vary as a function of the input networks, we compute a different $\alpha$ for each method and for each network pair. For instance, in the previous example of SANA-EC and \textit{RNorvegicus} and \textit{SPombe}, $\beta=0.5$ yields $\alpha=0.92$.

In Figure~\ref{fig:biogrid_beta0,5} we see the topological scores of the methods in inter-species alignments when setting $\beta=0.5$. SANA-LG, HubAlign and L-GRAAL get similar results, except in LCCS, where SANA-LG is generally worse (see discussion below). As expected, SANA-EC and SANA-S3 dominate in their respective objective functions.

\begin{figure}
\centering
\includegraphics[width=0.99\linewidth]{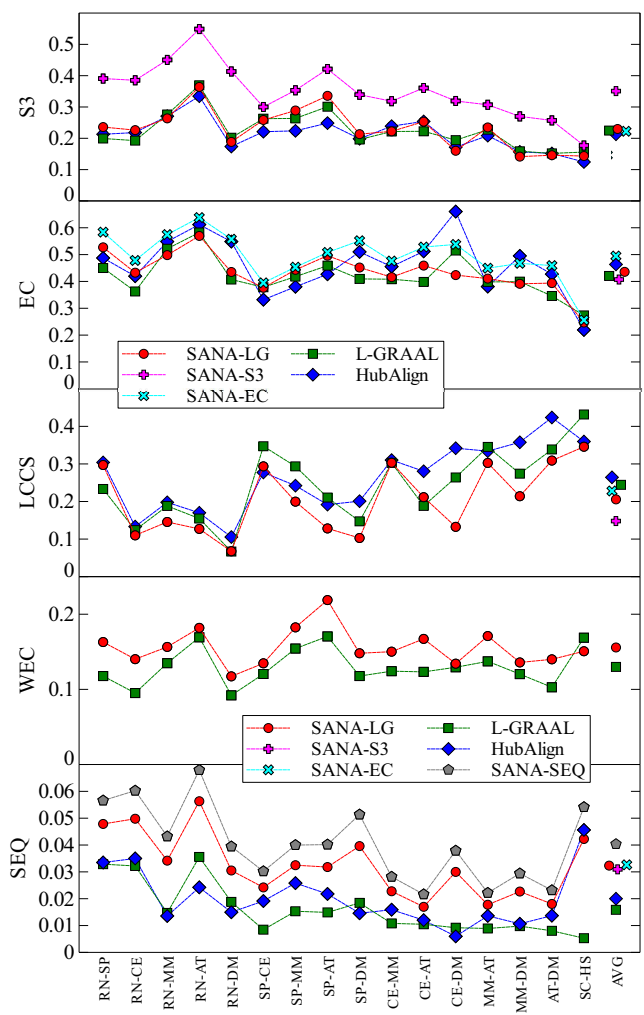}
\caption{Scores of the three topological target measures, S3, EC, and LCCS, in addition to graphlet-based WEC and Sequence (SEQ), for the BioGRID dataset when setting $\beta=0.5$. The WEC plot compares L-GRAAL's and SANA-LG's common topological objective function (it is analogous to Figure~\ref{fig:lgraalobjfuncomparison} but with $\beta=0.5$). For SANA-EC and SANA-S3 we only show the averages for clarity, except in their respective objective functions. In the SEQ plot, we have added SANA-SEQ, which optimizes Sequence alone (i.e., $\alpha=1$), to give a notion of the upper-bound of Sequence scores for SANA.}
\label{fig:biogrid_beta0,5}
\end{figure}

\begin{figure}
\centering
\includegraphics[width=0.99\linewidth]{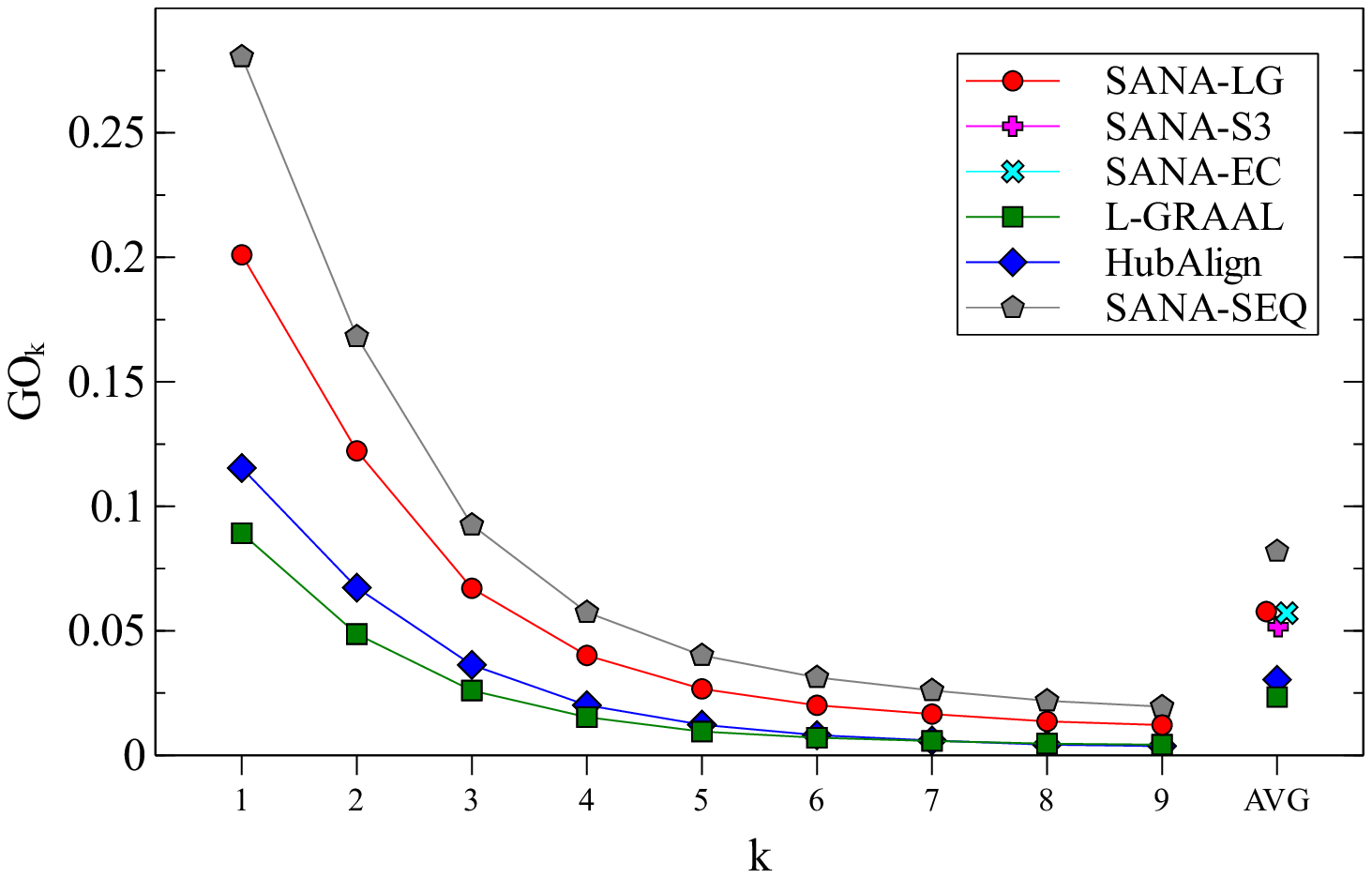}
\caption{Comparing the biological value of methods: average of each $\mbox{GO}_k$ scores among the 16 network pairs from the BioGRID dataset when using $\beta=0.5$. For example $k=1$ measures the fraction of aligned proteins that share 1 GO term; $k=9$ measures the fraction of aligned proteins that share 9 GO terms. The most egregiously common GO terms are eliminated since they add no useful information (see Section~\ref{eval}).}
\label{fig:biogridgo}
\end{figure}

Figure~\ref{fig:biogridgo} shows the biological quality of the alignments obtained by each method. All the variations of SANA outperform the other methods. Thus, SANA is capable of optimizing any topological measure better than the other methods while at the same time obtaining better biological results.

\subsection{Running time}

HubAlign, with a time complexity of $O(n^2\log n)$ is the fastest method. Its execution time for the largest pair of networks, \textit{SCerevisiae} and \textit{HSapiens}, was 33 minutes. In the case of SANA and L-GRAAL the maximum running time is a parameter specified by the user, which we set to one hour. However, in some instances L-GRAAL finished earlier than that. By design, SANA always uses the full allocated time.

\section{Discussion}

If the past 10 years of network alignment have proven anything, it is that the choice of which objective function(s) is (are) most important is unclear. EC was superseded by S3 because it was noted that EC is too simplistic because it has no incentive to match density fluctuations in the two networks; S3 was designed to explicitly favor alignments that match density fluctuations across the networks.  In some sense, EC was important early on because initial EC scores were so low (in the single digits of percentages) that any increase in EC was seen as a good thing.  As EC became saturated, it became clear that it was too simplistic and S3 replaced it.

We believe LCCS suffers an analogous shortcoming.  When LCCS values were low in the past, pushing them higher was a good thing.  But recent results suggest that LCCS values are close to being saturated. Eking out a larger LCCS is not hard today: one could, for example, perform a single ``swap'' operation (Figure~\ref{fig:operators}) that connects the two largest connected components of the common subgraphs with a single edge, thus creating a larger one, with minimal effect on all other measures. We even considered creating a version of SANA that explicitly optimizes LCCS to prove our point, but unfortunately LCCS cannot easily be incrementally calculated (\S \ref{implementation}), and so our hypothetical SANA-LCCS would take too long to run.  At this point, we believe that small differences in LCCS are best sacrificed in favor of better biological significance.  This is why we are not concerned that SANA-LG is not best in LCCS; we are happier that all versions of SANA produce better biological value than any other method while simultaneously producing comparable or better topological results in all measures except LCCS.

It is clear that when SANA explicitly optimizes a particular objective function (EC, S3, or WEC), it does so better than any existing method. What is also interesting is that SANA-LG matches or beats all other methods in these measures while {\em simultaneously} doing significantly better in biological measures (SEQ and $GO_k$). %This suggests that L-GRAAL's objective function, WEC, provides more biological significance than either EC or S3.

%We believe that, for now, WEC based on graphlets is the best topological measure. Five years ago \citep{GRAAL} it was shown that graphlets alone are capable of recovering significant biological information. SANA-LG's sequence scores (Figure~\ref{fig:biogrid_beta0,5}) and GO coverage (Figure~\ref{fig:biogridgo}) outperform all other methods, including SANA-EC and SANA-S3, while remaining competitive in all other topological measures against non-SANA competitors. %Note that in one case (SC-HS), L-GRAAL anomalously achieves a much better WEC score than SANA-LG. We note, however, that L-GRAAL also obtains anomalously \textit{low} biological scores in that case, as can be seen in the SEQ plot. This one case also scored badly in GO terms (not depicted). After verifying we used the correct input parameters, we are not sure why L-GRAAL anomalously seems to have focused on topology at the expense of biology in this one case.

While SANA-S3 grossly outperforms all other methods in optimizing S3, its sequence and GO scores are slightly lower than SANA-LG (although SANA-S3 still outperforms all non-SANA methods in sequence and GO measures). %This suggests again that graphlets provide the most comprehensive topological information capable of recovering biological information.

In conclusion, we have introduced a new simulated annealing search algorithm that can optimize any objective function when performing network alignment.  We have shown that SANA optimizes any explicit objective function better than any existing method, and produces solutions that score better in virtually every way compared to existing solutions, and does so faster than current methods. %We have also shown, so far as we know for the first time, that when compared on a level playing field, a graphlet-based topology measure produces better biological information than other topological scoring methods.

\section*{Acknowledgments}

We are very grateful to Tijana Milenkovi\'{c} and Vikram Saraph for directions on how to perform GO similarity analysis. We are also very grateful to No{\"e}l Malod-Dognin for assistance with sequence and GO similarity analysis on the BioGRID dataset. We thank Oleksii Kuchaiev for his insights on how to compute a statistically significant measure based on GO terms. We thank Yuki Sawa for help obtaining the sequence data of the Yeast and Human dataset.

\paragraph{Funding} The research has been supported by the Balsells fellowship.

\appendix

\section*{Appendix: EC is edge coverage, not edge ``correctness''}

When two pairs of aligned nodes both have an edge between them, we say that the edges are ``covered'' by each other.  Given that there are thousands of edges in both networks, this fact alone, for this particular pair, implies nothing ``correct'' about this particular alignment.  Furthermore, as Figure~\ref{fig:syntheticyeast} demonstrates, SAGA-LG is capable of maintaining virtually $100\%$ edge coverage even as the node correctness (and ``correctness'' here does have meaning) drops significantly.

In other words, there is nothing ``correct'' about edge correctness. In fact, if one is aligning two cliques, every alignment has $100\%$ EC, even if every node and edge is in the wrong place. The second author (W. Hayes) was in the room when this term was coined for this measure, and at the time it did not occur to anybody that it would be possible that EC could stay so high while the value of the alignment diminished so significantly. It is time to rename this measure to a more meaningful name: edges can be covered independently of their ``correctness'', and so the word ``correct'' is misleading. In fact, EC is so incorrect that S3 was invented to replace it. It is time to stop calling it ``correct''. We are doing a disservice to the long-term community of network alignment scientists by maintaining such a misleading term.

\bibliographystyle{natbib}
\bibliography{Bibliography}

\end{document}